\documentclass[aip,preprint]{revtex4-2}

\usepackage{amsmath}
\usepackage{graphicx}
\usepackage{subcaption}
\usepackage{xcolor}

\begin{document}


\title{Phonons of metallic hydrogen with quantum Monte Carlo}



\author{Kevin K. Ly}
\email[Author to whom correspondence should be addressed: ]{kkly2@illinois.edu}
\affiliation{Department of Physics, University of Illinois at Urbana-Champaign}

\author{David M. Ceperley}
\affiliation{Department of Physics, University of Illinois at Urbana-Champaign}


\date{\today}

\begin{abstract}
    We describe a simple scheme to perform phonon calculations with quantum Monte Carlo (QMC) methods, and demonstrate it on metallic hydrogen.
    Because of the energy and length scales of metallic hydrogen, and the statistical noise inherent to QMC methods, the conventional manner of calculating force constants is prohibitively expensive.
    We show that our alternate approach is nearly 100 times more efficient in resolving the force constants needed to calculate the phonon spectrum in the harmonic approximation.
    This requires only the calculation of atomic forces, as in the conventional approach, and otherwise little or no programmatic modifications.
\end{abstract}

\maketitle 

\section{Introduction}
\label{s:intro}
The normal modes of vibration in lattices, or phonons, play a significant role in the structural properties of solids.
For metallic hydrogen in particular, an accurate characterization of its lattice dynamics is desirable, because it is hypothesized to be a phonon-mediated superconductor \cite{superconductor}, and because experiments for dense hydrogen can only probe structural information indirectly through vibrational excitations.
Phonon calculations for metallic hydrogen have been carried out with density functional theory (DFT) \cite{metallic_hydrogen_phonons_2016, metallic_hydrogen_phonons_2020, PIMD_phonons}, but for dense hydrogen DFT can produce inconsistent results depending on the choice of exchange-correlation, and must be benchmarked against more accurate electronic methods like quantum Monte Carlo \cite{xc_benchmarks}.
It has been pointed out that while the electronic band structure can mostly be understood without explicit electron-electron interactions, the same is not true for the phonon band structure \cite{metallic_hydrogen_phonons_2016}.
With QMC methods, electronic correlations can be addressed directly.
To our knowledge, no phonon calculations for dense hydrogen have been performed using QMC methods.

Liu et al. \cite{fc_estimator} calculated the vibrational modes of small molecules by implementing an estimator for force constants.
However, this estimator formally has infinite variance, and has not been tested in larger systems.
Using atomic forces in a supercell, Nakano et al. \cite{VMC_phonons} performed a phonon calculation in diamond with variational Monte Carlo (VMC), enabled by an emphasis on reducing the noise in their calculations.
We take a similar approach, but because of the energy and length scales of metallic hydrogen, a prohibitively expensive reduction in noise would be required to resolve the phonon spectrum, especially if we want to incorporate more accurate and expensive QMC techniques such as diffusion or reptation quantum Monte Carlo.
We propose an alternative displacement protocol to accommodate calculations that do not require prohibitively high resolution.
This alternative protocol is simple, requires no modification of the underlying electronic structure method, and gives results equivalent to the conventional protocol.

In section \ref{s:harmonic_approx}, we review the harmonic approximation and the calculation of force constants from first principles.
In section \ref{s:random_protocol} we describe an alternative scheme for calculating these force constants, and explain why it is needed for metallic hydrogen.
The remainder of section \ref{s:methods} is dedicated to the details of the calculations.
In section \ref{s:dft_tests} we show that our alternative scheme works in practice; to see why it works in principle, see the appendix.
In section \ref{s:QMC_results} we present the QMC phonon band structures for metallic hydrogen at two different densities.
We also investigate the energy of two of these phonon modes, as a final check on our method.

\section{Methods}
\label{s:methods}

\subsection{The harmonic approximation}
\label{s:harmonic_approx}

In the harmonic approximation, the potential energy surface $U(\{\boldsymbol{R}\})$ is represented as a series about an equilibrium structure $\{\boldsymbol{R}^0\}$, truncated at second order
\begin{equation}
    \label{eq:harmonic}
    U(\{ \boldsymbol{R} \}) \approx U(\{ \boldsymbol{R}^0 \}) + \frac{1}{2} \sum_{I, J} (\boldsymbol{R}_I - \boldsymbol{R}_{I}^0) \frac{\partial^2 U}{\partial \boldsymbol{R}_I \partial \boldsymbol{R}_J} \bigg\vert_0 (\boldsymbol{R}_J - \boldsymbol{R}_{J}^0).
\end{equation}
This harmonic potential is exactly solvable and yields the well-known phonon solution \cite{ashcroft_and_mermin}.
In this paper we will not consider anharmonic effects, even though these are important for hydrogen \cite{metallic_hydrogen_phonons_2016, moving_protons, PIMD_phonons}.
To proceed, we must calculate the second derivatives of $U$, hereafter called force constants, since they are the first derivatives of the atomic forces.
In the simplest approach, these force constants are estimated by finite difference.
For example, using a first order forward difference, atom $I$ is moved by a small displacement $\Delta$ and one calculates
\begin{equation}
    \label{eq:finite_difference}
    \frac{\partial^2 U}{\partial \boldsymbol{R}_I \partial \boldsymbol{R}_J} \bigg\vert_0 = -\frac{\partial \boldsymbol{F}_J}{\partial \boldsymbol{R}_I} \bigg\vert_0 \approx -\frac{\boldsymbol{F}_J (\boldsymbol{R}_{I}^0 \to \boldsymbol{R}_{I}^0 + \Delta)}{|\Delta|}.
\end{equation}
where $\boldsymbol{F}_J$ is the force on atom $J$.
The number of such calculations can be minimized by the use of symmetries \cite{finite_difference, PHON, phonopy}.
Using an electronic structure method which calculates the forces on all of the atoms in a supercell, and a highly symmetric structure such as the Cs-IV or diamond structures, the force constants may be obtained in as few as one supercell calculation.

The above procedure can be generalized to a linear regression task \cite{phonons_by_fitting}, since in the harmonic approximation the forces are linear in the atomic displacements:
\begin{equation}
    \label{eq:linear_forces}
    \boldsymbol{F}_I = - \sum_J \frac{\partial^2 U}{\partial \boldsymbol{R}_I \partial \boldsymbol{R}_J} \bigg\vert_0 (\boldsymbol{R}_J - \boldsymbol{R}_{J}^0).
\end{equation}
In this way we can incorporate more than just the minimum number of force calculations, which may at first seem excessive but will actually be advantageous.
This generalization has been used in the temperature dependent effective potential approach by fitting force constants to \emph{ab initio} molecular dynamics trajectories \cite{TDEP}.
In our approach such a trajectory is not necessary, since we limit ourselves to the harmonic regime.
That is, the force constants we calculate are meant to be direct estimates of the second derivatives of $U$.

\subsection{Force constants by random sampling}
\label{s:random_protocol}

In order to calculate force constants, we perform $N_s$ force calculations, where $N_s$ is equal to or greater than the minimum number of calculations required by symmetry.
In a given calculation, \emph{every} atom is displaced from its site $\boldsymbol{R}_{I}^0 \to \boldsymbol{R}_{I}^0 + \boldsymbol{\chi}_I$ by a vector whose components are random numbers uniformly drawn from the interval $(-\Delta, \Delta)$.
The resulting atomic forces are generally larger in magnitude than they would be if only a single atomic coordinate had been displaced by $\Delta$.
In QMC, the resolution is essentially independent of how many atoms are displaced.
Consequently, the forces can be estimated with a much higher signal to noise ratio.
It can be shown (see appendix) that in the limit of perfect sampling ($N_s \to \infty$) and without any statistical noise, this approach estimates the second derivatives of $U$ with a leading order correction that scales as $\Delta^2$, comparable to a centered finite difference estimate.
In practice, finite sampling and the presence of statistical noise can be quantified and do not significantly hinder the procedure, as we will show.
The $\mathcal{O}(\Delta^2)$ behavior is maintained as long as we sample configurations with inversion symmetry: if we perform a calculation using displacements $\{\boldsymbol{\chi}\}$, we also include a calculation with the opposite displacements $\{-\boldsymbol{\chi}\}$.
Force constants are then estimated by linear regression over the $N_s$ samples.

If the random samples alone are not enough to resolve some of the lower frequency modes, where the errors due to noise tend to be greatest, then displacements along these modes can be included.
As in a frozen phonon calculation, the atoms are displaced according to the mode of interest.
The atomic forces are calculated and then included in the regression for the force constants.
Unlike a typical frozen phonon calculation, the mode of interest need not be known beforehand; it can be determined from an initial phonon calculation based on the random displacements just described.
From this initial calculation we can determine if these additional displacements are required to improve the resolution of certain modes, and the polarization of these modes will be available.
Since most (if not all) atoms are moved in a given phonon mode, the forces are still larger than in a single displacement calculation.
By construction, such displacements are meant to resolve particular phonon modes, whereas random displacements can be understood as superpositions of all modes and contain significantly more information.

\begin{figure}
    \begin{subfigure}{0.49\linewidth}
        \centering
        \includegraphics[width=\textwidth]{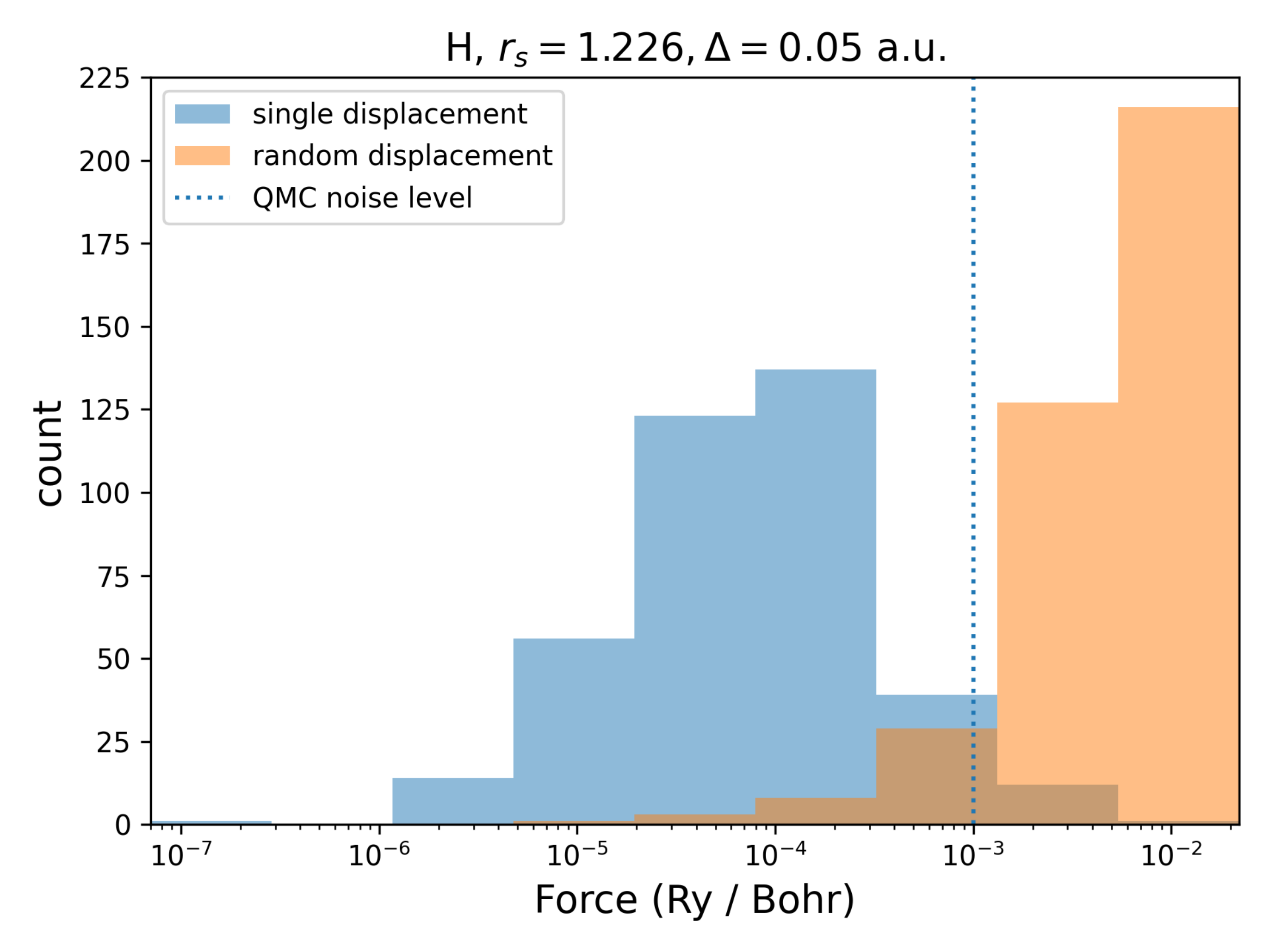}
        \caption{}
        \label{fig:histogram_h}
    \end{subfigure}
    \begin{subfigure}{0.49\linewidth}
        \centering
        \includegraphics[width=\textwidth]{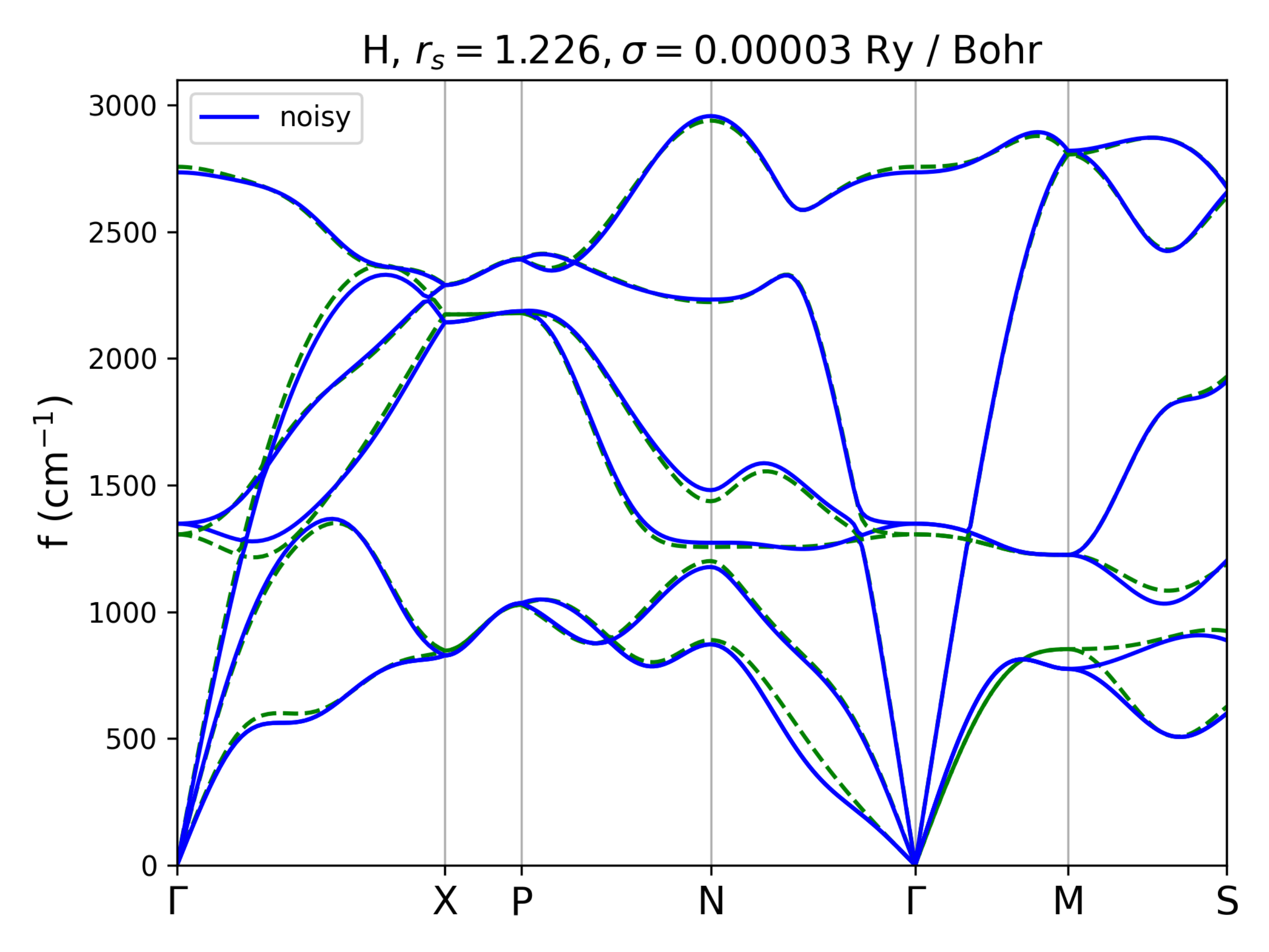}
        \caption{}
        \label{fig:perturbed_h}
    \end{subfigure}
    \caption{
        (a) Histogram of force magnitudes from a DFT single displacement calculation (blue) as well as a random displacement calculation (orange); the darker color is the overlap of the two.
        Note that the horizontal axes are on a log scale.
        A dashed vertical line indicating our QMC resolution is drawn for reference.
        (b) The forces from these single displacement calculations are used to calculate the phonon band structure, shown by the dashed green lines.
        The same forces, with the addition of simulated noise, are used to calculate a comparison phonon band structure, shown in blue.
    }
    \label{fig:signal_to_noise}
\end{figure}

The need for improved resolution is illustrated in Figure \ref{fig:signal_to_noise}.
Atomic forces were calculated with DFT PBE for a configuration in which only one of the atoms was displaced by 0.05 Bohr.
With reptation quantum Monte Carlo and a 128-atom supercell, a resolution of $10^{-3}$ Ry / Bohr is achievable with a calculation of reasonable length.
However, in Figure \ref{fig:histogram_h} we see that this resolution is not sufficient to resolve a majority of the forces in the single displacement calculation.
More explicitly, we simulated statistical noise on these forces by adding to each component a random number drawn from a Gaussian distribution with width $\sigma$.
In order to achieve the resolution seen in Figure \ref{fig:perturbed_h}, this noise must be limited to $10^{-5}$ Ry / Bohr, a factor of 100 smaller, which can be achieved in QMC by running 10000 times longer.

When all of the atoms are displaced according to the described random displacement protocol, many of the resulting forces are nearly 100 times greater.
If we maintain a resolution of $\sigma = 10^{-3}$ Ry / Bohr, the signal to noise ratio for the majority of the forces is 10 if all atoms are displaced, and 0.1 if only one is displaced.
The forces can also be made larger by choosing a larger finite difference $\Delta$, but note that 0.05 Bohr is already 3\% of the nearest neighbor spacing in this structure, and the forces increase only linearly with $\Delta$.

\subsection{QMC methods}
\label{s:qmc_methods}

In variational Monte Carlo (VMC) the properties of an explicit many-body trial wavefunction are calculated with Monte Carlo integration.
Given a trial wavefunction $\Psi_T$, the observable $A$ is computed as
\begin{equation}
    \label{eq:vmc}
    \langle A \rangle = \frac{ \int dR \ A(R) |\Psi_T (R)|^2 }{ \int dR \ |\Psi_T (R)|^2 },
\end{equation}
with specialized techniques for sampling many-electron wavefunctions.
The utility of VMC is its ability to accommodate sophisticated trial wavefunctions that explicitly encode many-body correlations.
Trial wavefunctions can be systematically improved using the variational principle.

In diffusion Monte Carlo (DMC), the ground state wavefunction $\Phi$ of a hamiltonian $H$ is obtained as the equilibrium solution of the imaginary-time Schr\"{o}dinger equation
\begin{equation}
    \label{eq:imaginary_time}
    -\frac{\partial \Phi}{\partial t} = (H - E_T) \Phi
\end{equation}
where $E_T$ is an adjustable constant used to calculate the ground state energy.
The evolution is achieved by simulating a diffusion and branching procedure \cite{DMC}.
For a many-electron system $\Phi$ cannot be strictly positive, and a straightforward adaptation of the diffusion procedure leads to the sign problem.
A common workaround is the fixed-node approximation \cite{fixed_node_introduction}: a trial wavefunction $\Psi_T$ is projected to the ground state using the diffusion procedure, but is prevented from changing its zeros, or nodes.
Consequently the true ground state $\Phi$ is only obtained if $\Psi_T$ has the same nodes as $\Phi$, otherwise the procedure gives an upper bound to the true ground state energy \cite{FN_upper_bound}.
For an observable $A$ which does not commute with the hamiltonian, like the force, DMC calculates the mixed estimate $\langle \Phi | A | \Psi_T \rangle$.
The pure estimate can be approximated by extrapolation \cite{extrapolation}
\begin{equation}
    \label{eq:extrapolation}
    \langle \Phi | A | \Phi \rangle \approx 2 \langle \Phi | A | \Psi_T \rangle - \langle \Psi_T | A | \Psi_T \rangle
\end{equation}
which relies on the VMC estimate for the second term.
Even with the fixed-node approximation, DMC is generally very accurate and less biased than methods like DFT \cite{DMC_benchmarks}, yielding lower energy and better forces than VMC.

In reptation quantum Monte Carlo (RQMC), a trial wavefunction $\Psi_T$ is projected to the ground state $\Phi \approx \exp( -\tau H ) \Psi_T$ by exploiting the path-integral formulation of the Schr\"{o}dinger equation.
Instead of having random walkers diffusing and branching, they are assembled into a ``reptile'' which advances in a snake-like manner to sample paths \cite{RMC}.
For a many-electron system the fixed-node (or fixed-phase for complex trial functions) approximation is used.
The advantage of RQMC over DMC is that it directly calculates the pure estimate $\langle \Phi | A | \Phi \rangle$, subject to the fixed-node approximation.

Forces in QMC are calculated using a modified Hellman-Feynman estimator \cite{force_estimator}.
Using Monte Carlo sampling the bare estimator has infinite variance because of the finite electronic density at the ion positions.
As such, it is replaced by a fitted estimator whenever electrons get close to an ion.
Because it relies on the Hellman-Feynman theorem, the wavefunction must either be at a variational minimum, or must not explicitly vary with the atomic coordinates, otherwise there will be a discrepancy between the force estimate and corresponding derivative of the energy.
Our trial wavefunctions do not follow these requirements strictly, and hence we find discrepancies of up to 10 mRy / Bohr in VMC.
Note from Figure \ref{fig:histogram_h} that this is comparable to the magnitude of the forces in these calculations, so these discrepancies are not minor.
However, in DMC and RQMC, the energy depends only on the nodal surface of the trial wavefunctions, so that such discrepancies are less severe.
Because RQMC directly calculates the pure estimate, this minimizes the reliance on our trial wavefunctions, so it will be our method of choice for our force calculations.
However, we find that our extrapolated DMC forces and RQMC forces are consistent within error bars.

\subsection{The Cs-IV structure}
\label{s:structure}

We focus on the candidate Cs-IV structure of metallic hydrogen \cite{cs_iv_structure}.
This is a highly symmetric structure, and consequently the force constants can be obtained in principle from a single force calculation.
We study two different densities, $r_s = 1.226$ and $r_s = 1.3$, where $r_s$ is the mean electronic spacing.
The corresponding pressures, as calculated with DFT PBE, are approximately 500 GPa and 300 GPa, respectively.

\begin{figure}
    \begin{subfigure}{0.49\linewidth}
        \centering
        \includegraphics[width=\textwidth]{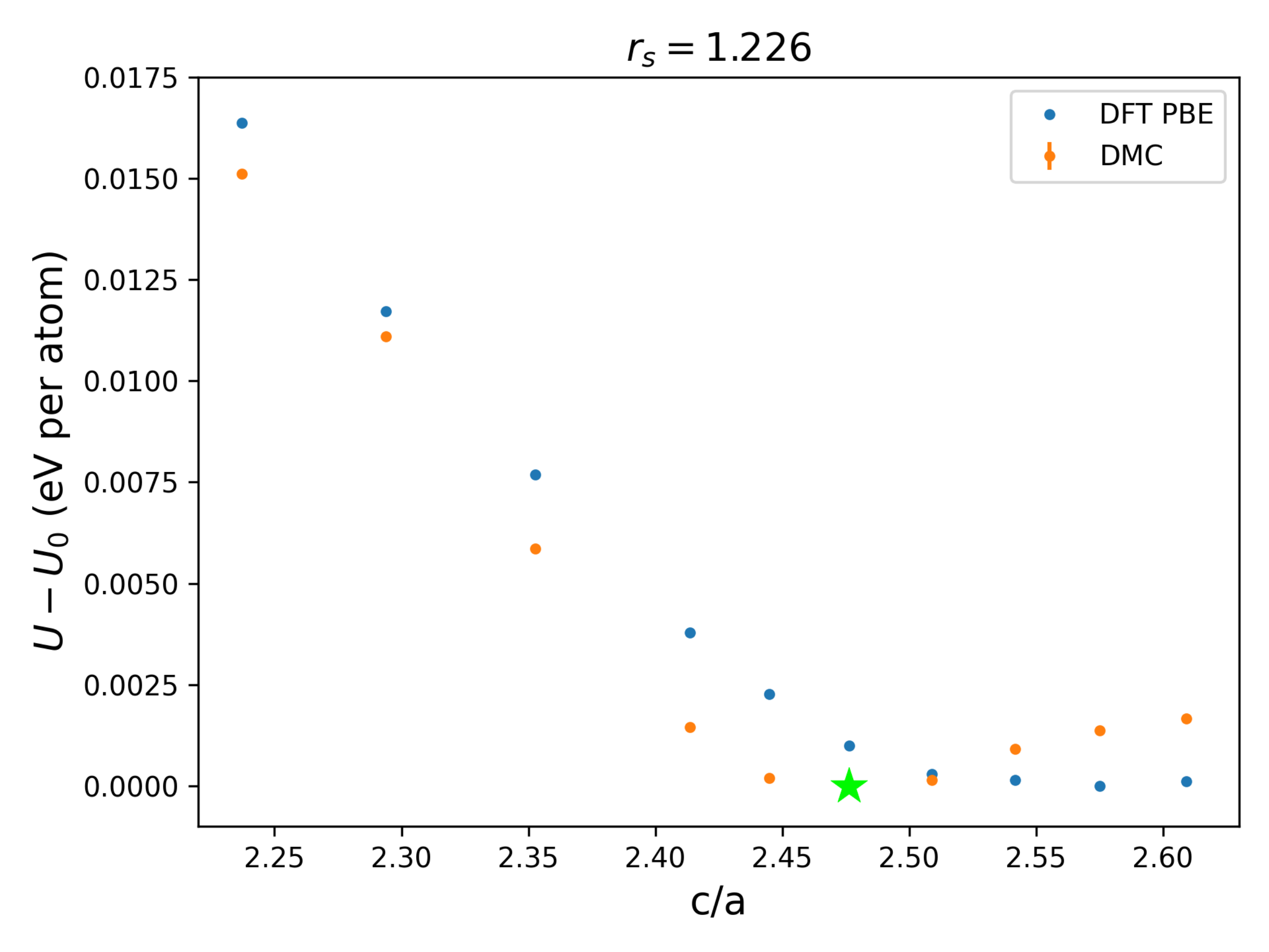}
        \caption{}
    \end{subfigure}
    \begin{subfigure}{0.49\linewidth}
        \centering
        \includegraphics[width=\textwidth]{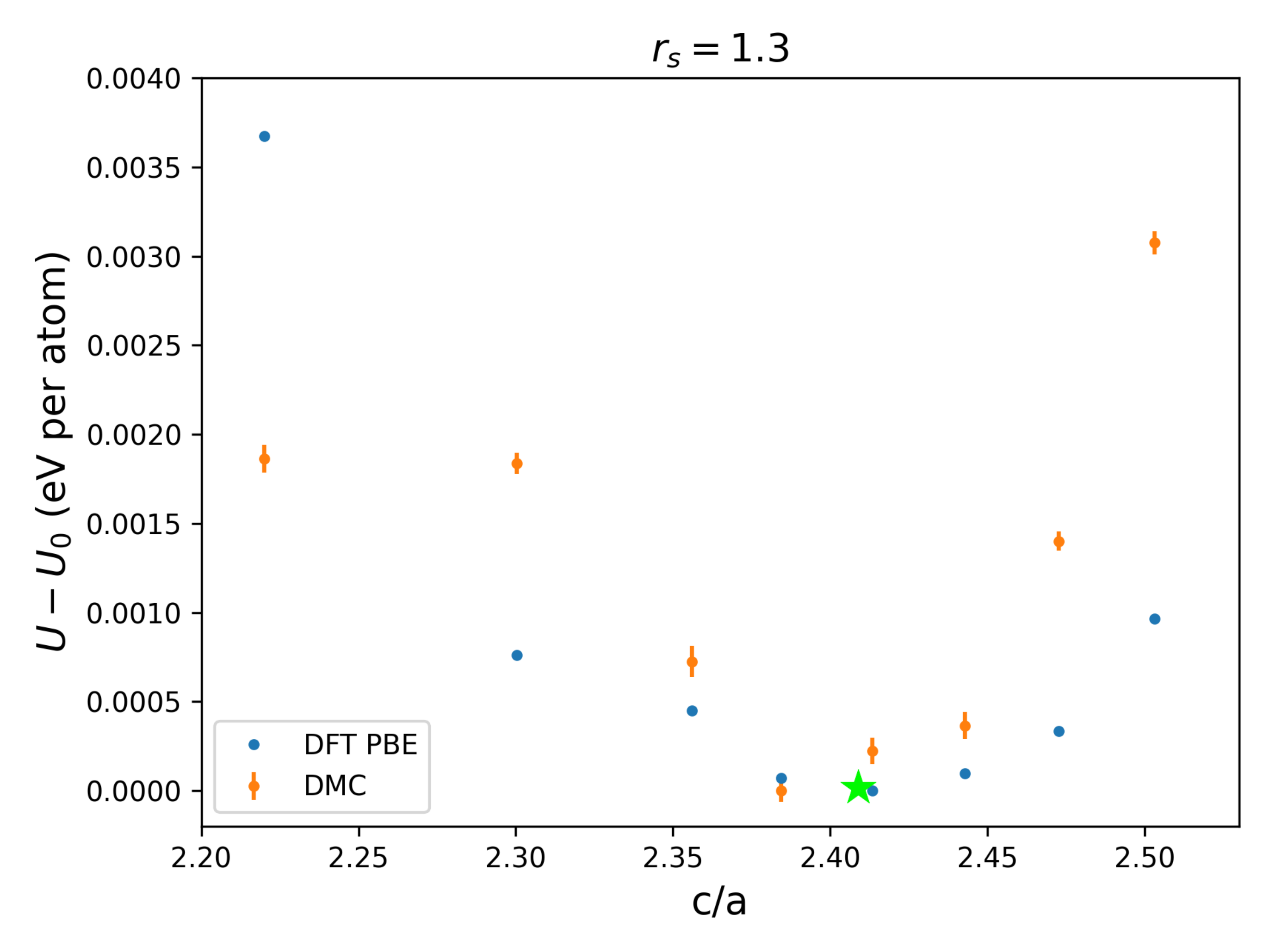}
        \caption{}
    \end{subfigure}
    \caption{
        DMC and DFT PBE energy as the $c/a$ ratio is adjusted, at (a) $r_s = 1.226$ and (b) $r_s = 1.3$.
        The $c/a$ ratios that are used for the phonon calculations are marked by the green star.
    }
    \label{fig:ca}
\end{figure}

At a given density, this structure has only one free parameter, the $c/a$ ratio.
For both densities, we calculated the energy for various $c/a$ ratios with DMC to search for the minimum.
The results are shown in Figure \ref{fig:ca}.
Note that these calculations are performed on a 128-atom supercell, and can be subject to considerable finite size effects \cite{finite_size_effects}.
For consistency, all calculations in this paper will use the same number of atoms.
The lattice parameters used are: $a = 2.32$ Bohr and $c/a = 2.475$ for $r_s = 1.226$, $a = 2.48$ Bohr and $c/a = 2.41$ for $r_s = 1.3$.

\subsection{Calculations}
\label{s:calcs}

DFT calculations are performed with \texttt{QUANTUM ESPRESSO} \cite{QE_2009, QE_2017} using a 128-atom supercell, a $4 \times 4 \times 2$ tiling of the 4-atom conventional cell.
We use a Troullier-Martins norm-conserving pseudopotential with a cutoff of 0.65 Bohr, a plane-wave kinetic energy cutoff of 300 Ry, and BZ integration over a shifted $7 \times 7 \times 7$ grid.
The PBE \cite{PBE} and PZ \cite{PZ} functionals are used, the former for its popularity in previous metallic hydrogen work, and the latter because the resultant orbitals perform slightly better in QMC, having lower variance.

VMC, DMC, and RQMC calculations are performed with \texttt{QMCPACK} \cite{QMCPACK}.
These are all-electron calculations; there is no pseudopotential.
Twist-averaging \cite{twist_averaging} is performed using a shifted $7^3$ grid.
We use Slater-Jastrow trial wavefunctions, using orbitals from DFT PZ calculations, and a Jastrow factor composed of both real-space and reciprocal space two-body terms.
While the orbitals change for each random displacement, the Jastrow factor is optimized at the undisturbed crystal structure and used in subsequent calculations.
As we described in section \ref{s:qmc_methods}, force calculations are done with RQMC.
We find that our RQMC calculations are converged with respect to projection time using a timestep of 0.01 Ha and a reptile with 300 beads.
Additional details can be found in the inputs provided in the supplemental material.

Force constants and band structures are calculated using \texttt{ALAMODE} \cite{ALAMODE}, which consists of two components: \texttt{ALM}, which calculates force constants, and \texttt{ANPHON}, which processes the force constants.
The symmetry analysis to minimize the number of force calculations required is handled internally by \texttt{Spglib} \cite{spglib}.
Linear regression is then performed using ordinary least-squares, for which the random displacement protocol is most easily understood (see appendix).
The resultant force constants in the supercell are then used to calculate the dynamical matrix in the primitive cell, from which the band structure can be obtained.
The eigenvectors and eigenvalues of these force constants are used for additional frozen phonon calculations.

\section{Results}
\label{s:results}

\subsection{Tests with DFT}
\label{s:dft_tests}

We first demonstrate that the random displacement method correctly estimates force constants, by checking it against a centered finite difference scheme.
There are three primary sources of error: statistical uncertainty from randomly sampling displacements, statistical uncertainty inherent to QMC calculations, and the $\mathcal{O}(\Delta^2)$ error from the higher order terms neglected in \eqref{eq:linear_forces}.
We can quantify the former two sources with a jackknife method \cite{jackknife}.
For example, as applied to the calculation of phonon band structures $\omega_n (\boldsymbol{k})$, we do the following:
\begin{itemize}
    \item Calculate force constants by fitting to all available force calculations, from which the final estimate of the band structure $\omega_n (\boldsymbol{k})$ is obtained.
    \item Remove force calculation $i$ from the data and fit force constants to this slightly smaller data set, from which a generally different estimate of the band structure $\omega_{n}^i (\boldsymbol{k})$ is obtained.
    \item Repeat the above step for all force calculations $i$, so that the uncertainty is estimated as
        \begin{equation}
            \label{eq:jackknife}
            \sigma_n (\boldsymbol{k}) \approx \sqrt{ \sum_i ( \omega_{n}^i (\boldsymbol{k}) - \omega_n (\boldsymbol{k}) )^2 }.
        \end{equation}
\end{itemize}
For a harmonic potential, which one would expect as $\Delta \to 0$, and no noise in the force data, every fit would yield the same result and the uncertainty would be 0.

\begin{figure}
    \begin{subfigure}{0.49\linewidth}
        \centering
        \includegraphics[width=\textwidth]{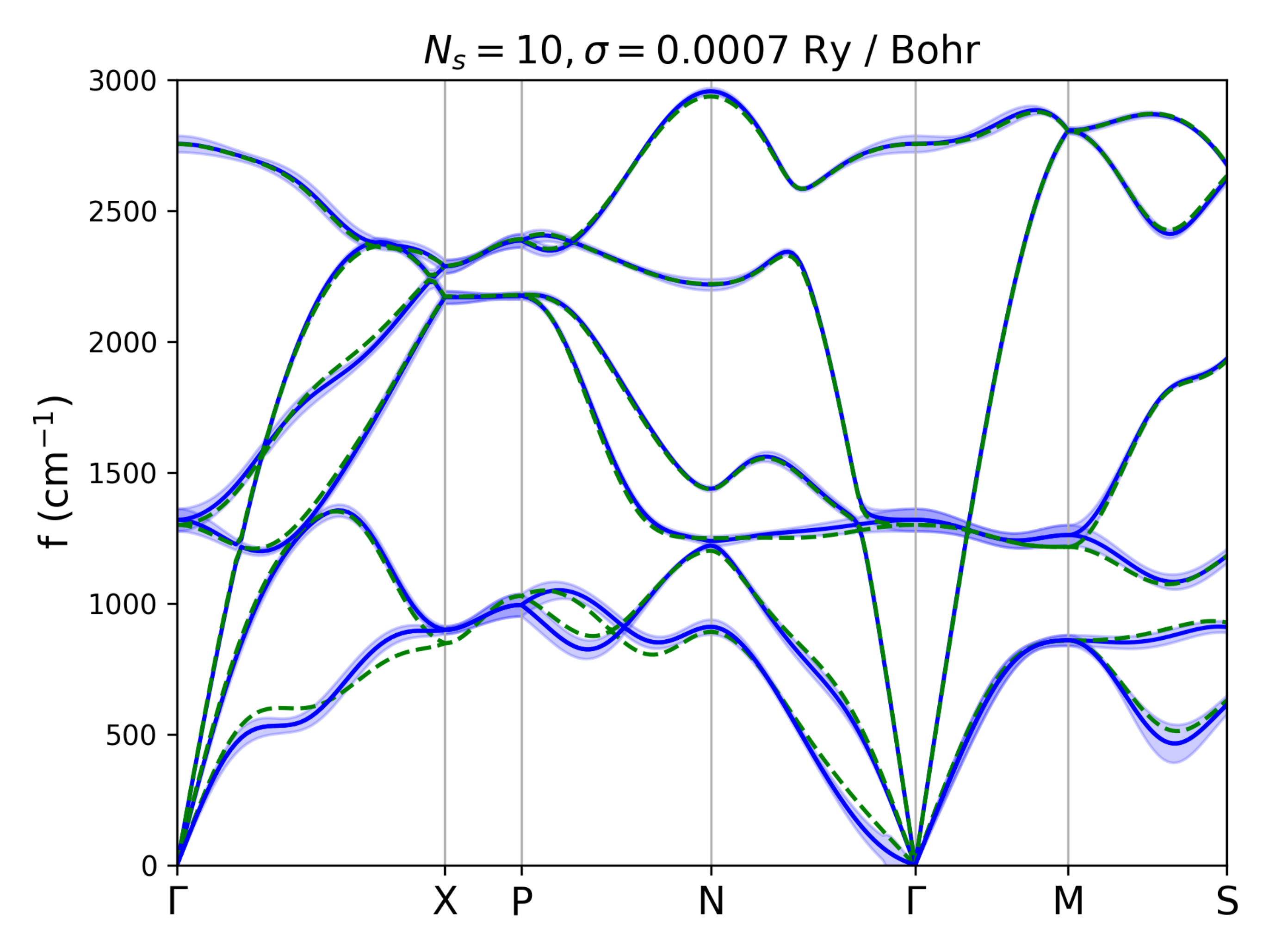}
        \caption{}
    \end{subfigure}
    \begin{subfigure}{0.49\linewidth}
        \centering
        \includegraphics[width=\textwidth]{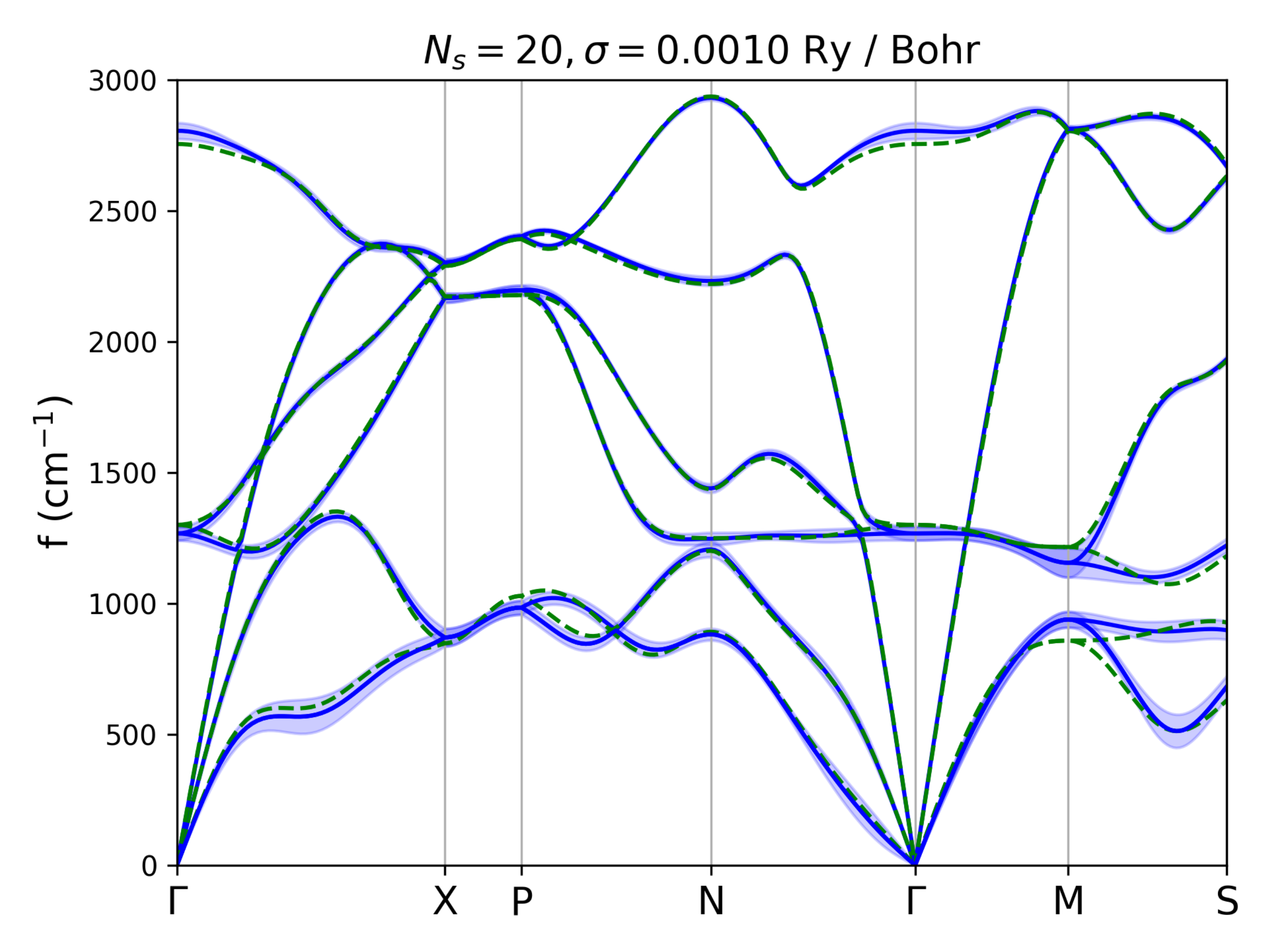}
        \caption{}
    \end{subfigure}

    \begin{subfigure}{0.49\linewidth}
        \centering
        \includegraphics[width=\textwidth]{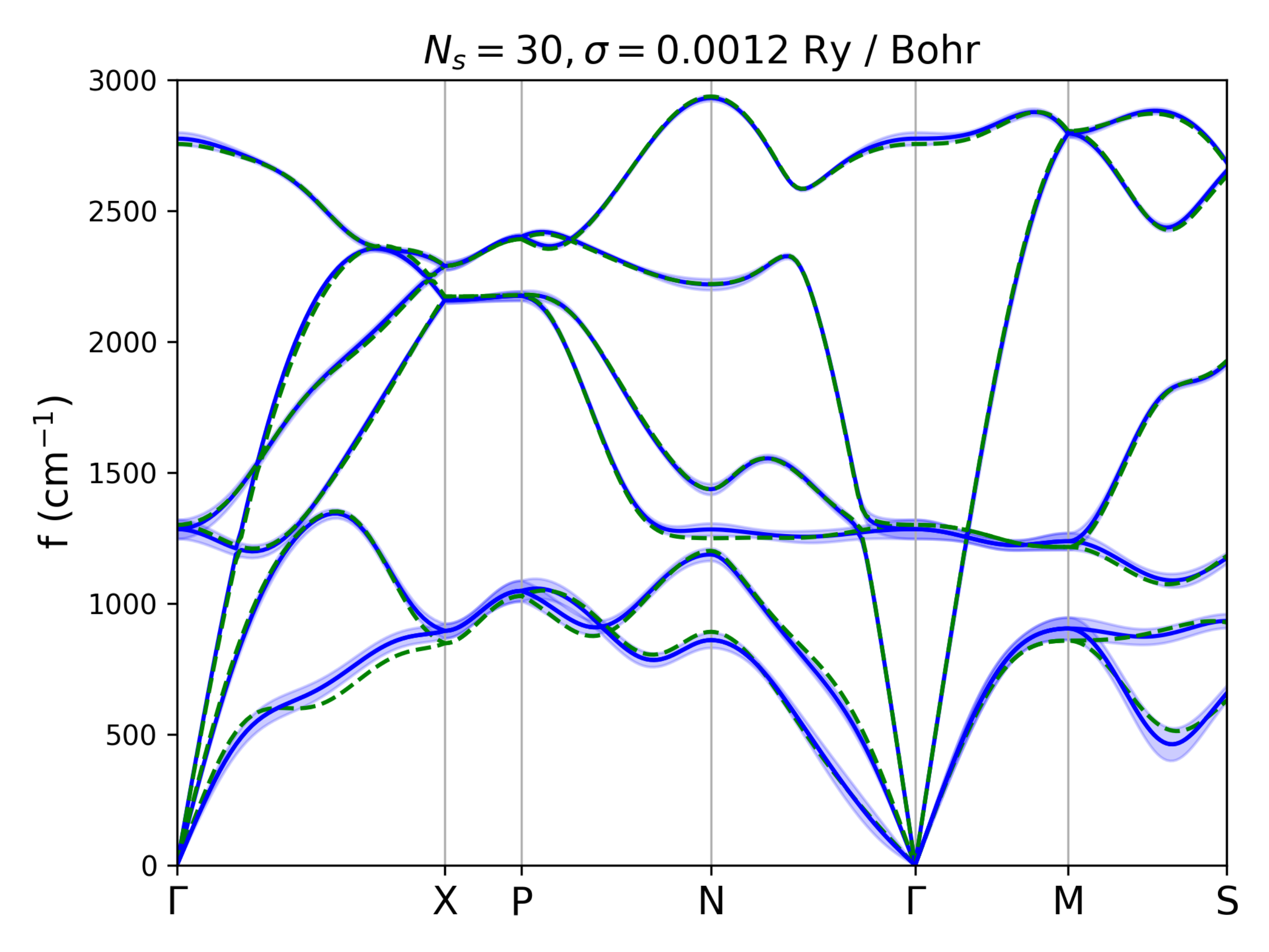}
        \caption{}
    \end{subfigure}
    \begin{subfigure}{0.49\linewidth}
        \centering
        \includegraphics[width=\textwidth]{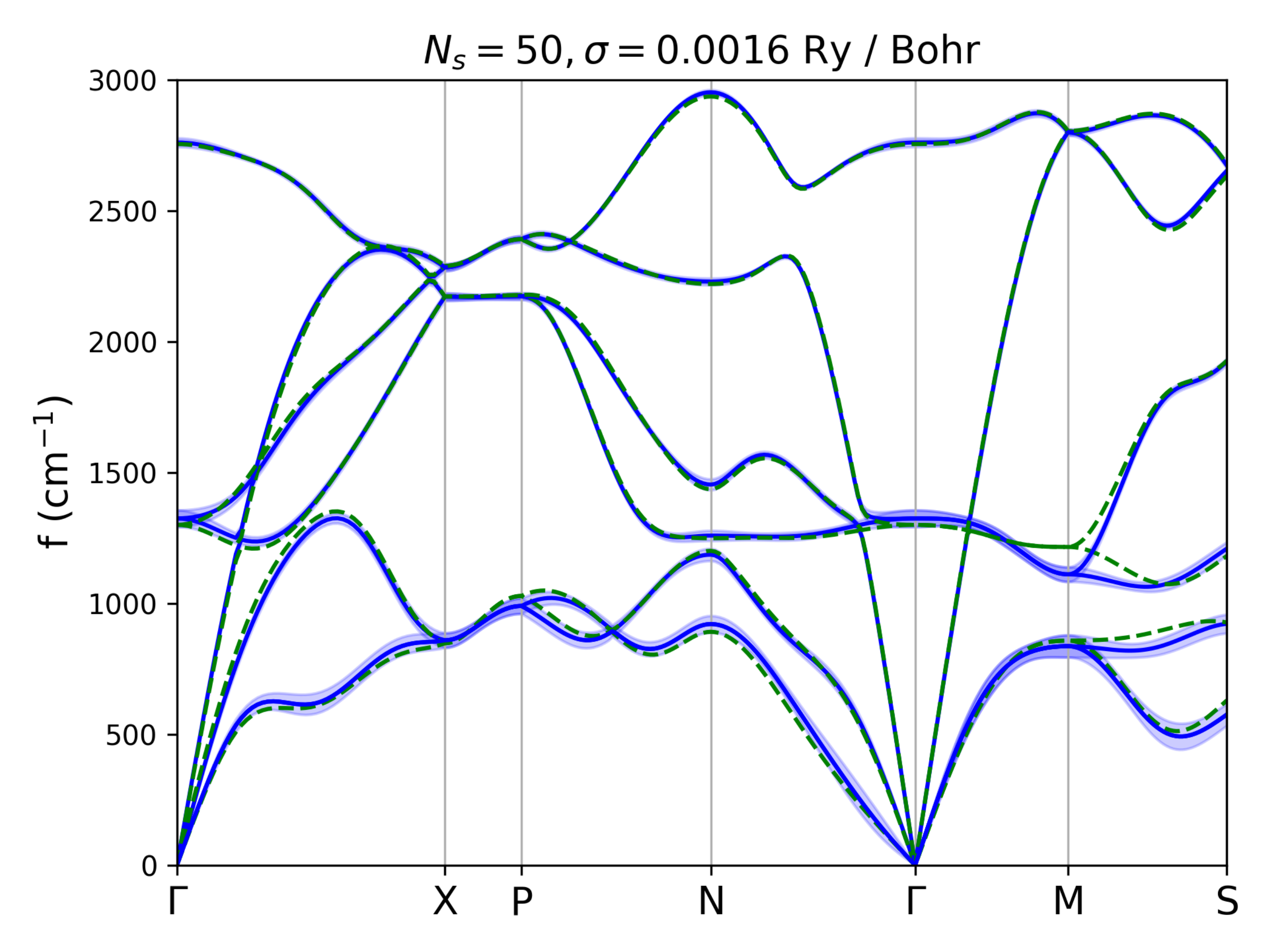}
        \caption{}
    \end{subfigure}
    \caption{
        Band structures calculated using the random displacement protocol (blue) with DFT PBE and $\Delta = 0.05$ Bohr, using (a) 10, (b) 20, (c) 30, and (d) 50 random samples and varying levels of simulated noise $\sigma$.
        For reference, a band structure is calculated using a centered finite difference scheme (green) with $\Delta = 0.007$ Bohr.
        The green curves are identical in all four plots.
        The statistical uncertainties (blue shade) are estimated by the jackknife method.
    }
    \label{fig:tests}
\end{figure}

We perform tests with DFT PBE at $r_s = 1.226$ to show that, in practice, our approach has acceptable resolution.
Shown in Figure \ref{fig:tests} are phonon band structures as calculated by the random displacement method for various numbers of random samples $N_s$ and various levels of simulated noise $\sigma$.
The level of noise is adjusted to make the calculations more comparable in expense.
For example, 10 QMC calculations is comparable to doing 50 QMC calculations if the 10 calculations are run 5 times as long.
Consequently, the statistical uncertainty in these 10 calculations is reduced by a factor of $\sqrt{5}$.
For the random displacements we choose $\Delta = 0.05$ Bohr as before, and we compare the results to the band structure obtained by a centered finite difference scheme with considerably smaller $\Delta = 0.007$ Bohr to demonstrate convergence.
While all of the tests appear to satisfactorily reproduce the conventional result, we favor the largest number of samples $N_s$ because our analysis of systematic errors is based on $N_s \to \infty$.

For the tests in which $N_s < 50$, a subset of $N_s$ force calculations is chosen such that if random displacement $\{ \boldsymbol{\chi} \}$ is included, then $\{ -\boldsymbol{\chi} \}$ is also included.
The tests are reproducible in the sense that they do not change qualitatively when a different random seed is used for simulating noise, or if a different subset of samples is chosen.
Based on these results, we conclude that, with a resolution of $\sigma = 0.002$ Ry / Bohr, $N_s = 70$ calculations should be sufficient to resolve the band structure.
Recall that the more conventional single displacement protocol would require a calculation nearly 10000 times as long.

\subsection{QMC phonons}
\label{s:QMC_results}

\begin{figure}
    \begin{subfigure}{0.49\linewidth}
        \centering
        \includegraphics[width=\textwidth]{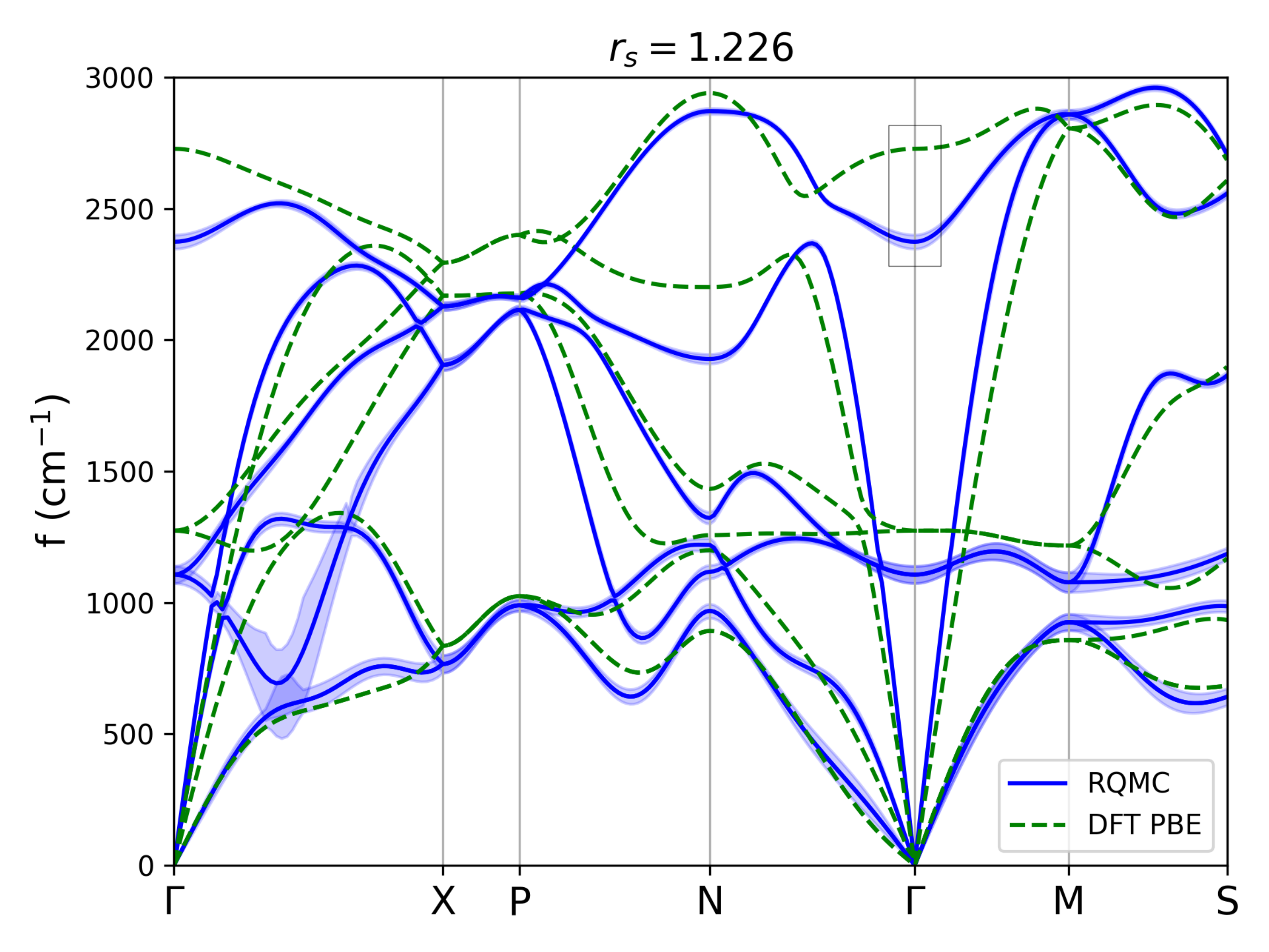}
        \caption{}
    \end{subfigure}
    \begin{subfigure}{0.49\linewidth}
        \centering
        \includegraphics[width=\textwidth]{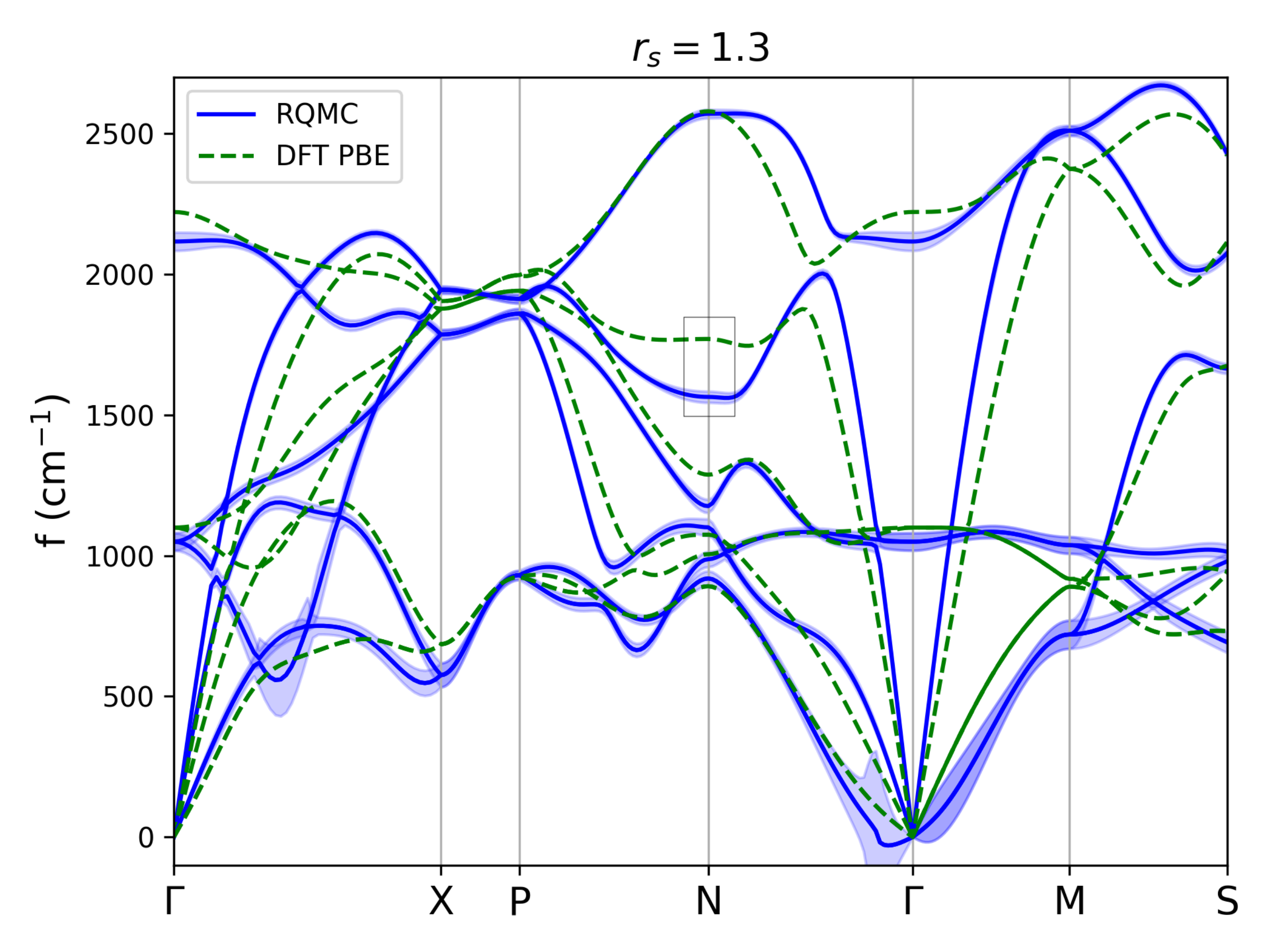}
        \caption{}
    \end{subfigure}
    \caption{
        Phonon band structures calculated using the random displacement method and RQMC forces (blue), alongside DFT (green), for (a) $r_s = 1.226$ and (b) $r_s = 1.3$.
    The highest mode at $\Gamma$ for $r_s = 1.226$ and second highest mode at $N$ for $r_s = 1.3$ that are investigated in Figure \ref{fig:frozen_phonons} are highlighted by the boxes.
    }
    \label{fig:QMC}
\end{figure}

Shown in Figure \ref{fig:QMC} are the phonon band structures as calculated with RQMC.
For the majority of modes, the resolution achieved is consistent with what we expected based on our tests from \S \ref{s:dft_tests}, with the exception of the lowest energy mode between $\Gamma X$.
For both densities, we included displacements along the lowest energy mode halfway along $\Gamma X$ in fitting.
Without them, the random displacement calculation yielded frequencies that were too low for these modes, for example resulting in an imaginary mode (with high uncertainty) for $r_s = 1.226$.
The inclusion of these frozen phonon calculations corrects these low frequencies, though they are still difficult to resolve, as indicated by the relatively large uncertainties.
For $r_s = 1.3$, the modes near $\Gamma$ along $N \Gamma$ and $\Gamma M$ also appear to be poorly resolved, with the band even becoming slightly imaginary.
However, this is an artifact of Fourier interpolation, since the modes between $N$ and $\Gamma$ are not actually commensurate with the supercell.

To see if our force calculations are consistent with our energies, we perform additional frozen phonon calculations that are not included in the fit.
For a mode with eigenvector (polarization) $\epsilon$ and eigenvalue $\lambda$, displacing the atoms according to this polarization with amplitude $x$ will result in the energy
\begin{equation}
    \label{eq:frozen_phonon}
    U \approx U_0 + \frac{1}{2} \lambda x^2.
\end{equation}
For a given mode, we calculate the energy for various ampltides $x$ and perform a fit to produce an independent estimate of the mode's eigenvalue.
The eigenvalue is related to the mode frequency $\omega$ by $M \omega^2 = \lambda$.

\begin{figure}
    \begin{subfigure}{0.49\linewidth}
        \centering
        \includegraphics[width=\textwidth]{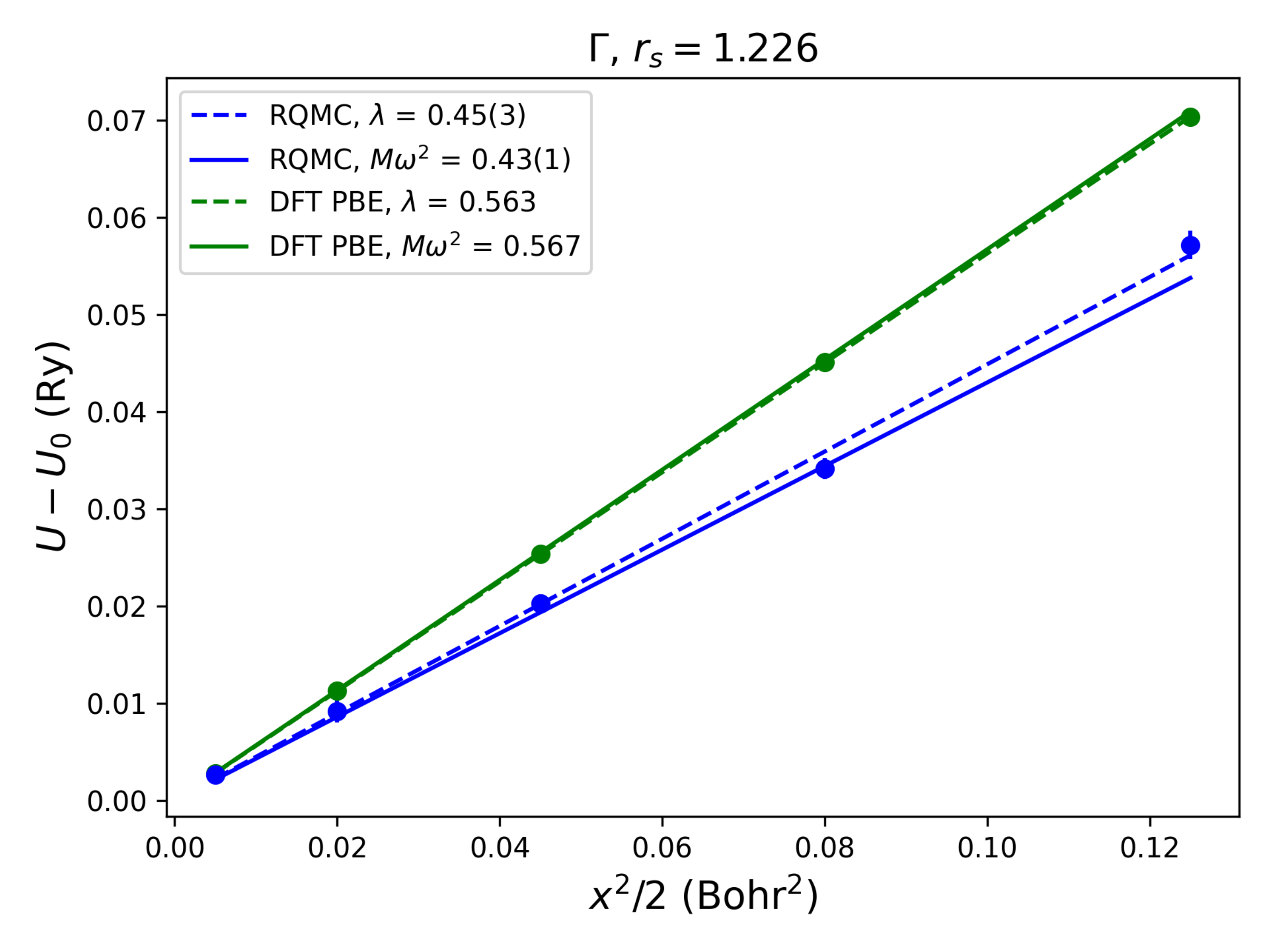}
        \caption{}
    \end{subfigure}
    \begin{subfigure}{0.49\linewidth}
        \centering
        \includegraphics[width=\textwidth]{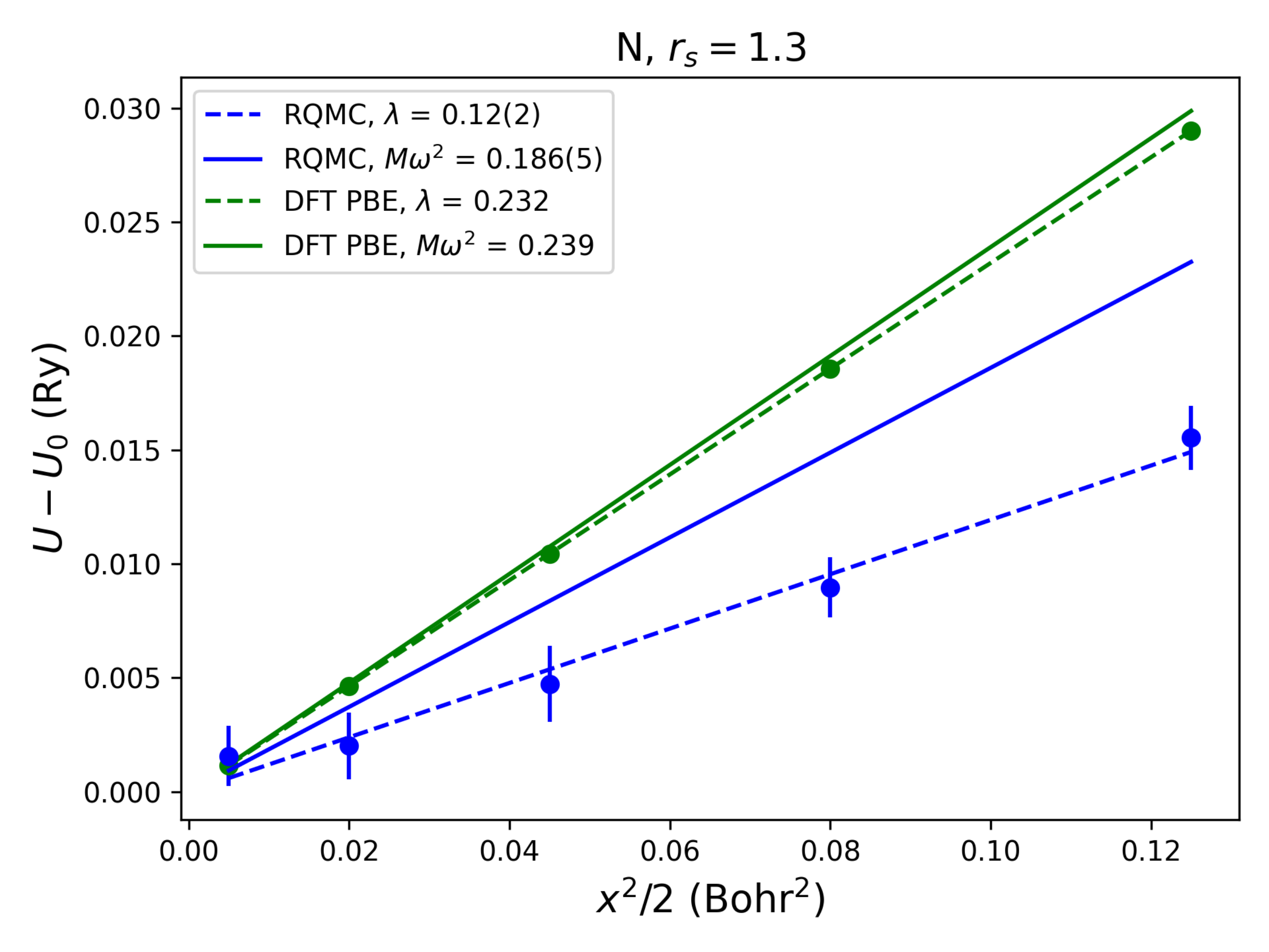}
        \caption{}
        \label{fig:N_mode}
    \end{subfigure}
    \caption{
        Energy as the atoms are displaced along (a) the highest mode at $\Gamma$ for $r_s = 1.226$ and (b) the second highest mode at $N$ for $r_s = 1.3$ with amplitude $x$.
        $\lambda$ is the slope of the linear fit, and $M \omega^2$ is calculated using the frequency from the phonon calculation.
        These are the two modes highlighted in Figure \ref{fig:QMC}.
    }
    \label{fig:frozen_phonons}
\end{figure}

We examine the highest mode at $\Gamma$ for $r_s = 1.226$ and the second highest mode at $N$ for $r_s = 1.3$, for which there is an apparent difference between RQMC and DFT.
Shown in Figure \ref{fig:frozen_phonons} are the energies of these modes plotted against $x^2 / 2$.
From these energies the eigenvalue $\lambda$ was estimated with a linear fit, and compared to the eigenvalue obtained from the phonon calculation $M \omega^2$.
If the forces correctly describe the derivatives of the energy, and if $x$ is small enough so that the energy is harmonic, these two eigenvalue estimates should be consistent.
This is the case for the DFT calculations, and the slight deviation can be attributed to anharmonic contributions.
This is also the case for RQMC in the $\Gamma$ mode, but not so in the $N$ mode.
Because the conditions of the Hellman-Feynman theorem are not exactly fulfilled, the force estimator has a bias which, although small, can be apparent as the energy scales get small.
We find that this bias tends to make the magnitudes of the forces larger, so that the resultant phonon calculations tend to slightly overestimate frequencies.
Nevertheless, these calculations still correctly show that the energies of these modes are clearly flatter in RQMC than in DFT.

From Figure \ref{fig:N_mode} we can also see why lower energy modes tend to be more difficult to resolve: one has to resolve smaller changes in energy, but the statistical noise is more apparent at these scales.
The presence of lower energy modes also accounts for the fact that the resolution seen in Figure \ref{fig:QMC} is not always as good as we expected from our DFT tests.

\section{Conclusion}

We described a strategy for calculating harmonic force constants in the presence of statistical noise.
This enabled us to calculate the phonons of metallic hydrogen with RQMC, where a more conventional method would have been nearly 100 times more expensive.
The strategy is simple and requires little additional development: none of the software described in this work needed any programmatic modification.
Coupled with progress in QMC and QMC force calculations, we believe this could be useful for enabling more challenging studies of lattice dynamics.

Because phonon modes in the harmonic approximation are highly constrained by symmetry, the band structures as calculated by QMC and DFT share many qualitative features.
However, in some modes, DFT appears to overestimate the frequency, meaning it overestimates the curvature of the energy surface.
We have shown this explicitly for two particular modes with frozen phonon calculations.
This could impact calculations of the superconducting transition $T_c$, where the highest frequency branches appear to have the greatest influence \cite{mode_specific}.
That the energy surface should be flatter in certain directions could impact the dynamics of metallic hydrogen, since the protons can be highly diffusive.


%
%

%

\section{Data availability}

The data that supports the findings of this study are available within the supplementary material.

\section{Supplementary material}

The supplementary material contains inputs describing the random displacements used, the details of the QMC calculations, the final QMC forces used, and the details of the phonon calculations.

\begin{acknowledgments}
    We thank Yubo Yang and Raymond Clay for their insights on \texttt{QMCPACK} and QMC forces.
    This research was suppored by grant DOE DE-SC0020177.
    This research is part of the Blue Waters sustained-petascale computing project, which is supported by the National Science Foundation (awards OCI-0725070 and ACI-1238993) the State of Illinois, and as of December, 2019, the National Geospatial-Intelligence Agency.
    Blue Waters is a joint effort of the University of Illinois at Urbana-Champaign and its National Center for Supercomputing Applications.
    This work made use of the Illinois Campus Cluster, a computing resource that is operated by the Illinois Campus Cluster Program (ICCP) in conjunction with the National Center for Supercomputing Applications (NCSA) and which is supported by funds from the University of Illinois at Urbana-Champaign.
\end{acknowledgments}

\bibliography{bibliography}

\appendix

    \section{Systematic error of the random sampling method}
    \label{s:appendix}

    Consider a function $U(x_1, \dots, x_{3N})$ of $3N$ coordinates, in our case representing the energy as a function of all of the coordinates in an $N$-atom supercell.
    For simplicity suppose the coordinates are shifted such that the equilibrium structure is given by $x_i = 0$ for all $i$.
    Expanding $U$ about this equilibrium and including terms beyond the harmonic ones, we have
    \begin{equation}
    \begin{split}
        U &= U_0 + \frac{1}{2} \sum_{i j} \Phi_{i j} x_i x_j + \frac{1}{6} \sum_{i j k} \Gamma_{i j k} x_i x_j x_k + \frac{1}{24} \sum_{i j k l} \Lambda_{i j k l} x_i x_j x_k x_l\\
        \implies F_{\alpha} &= -\sum_j \Phi_{\alpha j} x_j - \frac{1}{2} \sum_{j k} \Gamma_{\alpha j k} x_j x_k - \frac{1}{6} \sum_{j k l} \Lambda_{\alpha j k l} x_j x_k x_l\\
        \text{where} \quad \Phi_{i j} &\equiv \frac{\partial^2 U}{\partial x_i \partial x_j}\bigg\vert_{0},\ \Gamma_{i j k} \equiv \frac{\partial^3 U}{\partial x_i \partial x_j \partial x_k}\bigg\vert_{0},\ \Lambda_{i j k l} \equiv \frac{\partial^4 U}{\partial x_i \partial x_j \partial x_k \partial x_l}\bigg\vert_{0},
    \end{split}
    \end{equation}
    and $F_\alpha$ is the force in direction $\alpha$, obtained by taking the derivative of $U$ and changing the sign.
    Because they are derivatives, $\Phi, \Gamma, \Lambda$ are invariant under the permutation of their indices.
    Including more terms in the expansion for the following analysis is straightforward.

    For a given input $\{ x_i \}$, our electronic structure methods will yield the corresponding energy $U$ and forces $F$, but we do not know their functional forms otherwise, i.e. we do not know what $\Phi, \Gamma, \Delta$ are.
    In the harmonic approximation we wish to know $\Phi_{\alpha \beta}$.
    A centered finite difference estimate in which we calculate the force on $\alpha$ after displacing $\beta$ by some small amount $\Delta$, with all other $x_i = 0$, yields
    \begin{equation}
    \begin{split}
        -\Phi^{e}_{\alpha \beta} &= \frac{F_{\alpha} (x_{\beta} = \Delta) - F_{\alpha} (x_{\beta} = -\Delta)}{2 \Delta}\\
        &= -\Phi_{\alpha \beta} - \frac{1}{6} \Lambda_{\alpha \beta \beta \beta} \Delta^2
    \end{split}
    \end{equation}
    where the superscript is used to distinguish our estimate from the true second derivatives.
    This is the familiar $\mathcal{O}(\Delta^2)$ error we expect from a centered finite difference estimate.

    The random displacement protocol we described can be thought of as uniformly sampling the region $x_1, \dots, x_{3N} \in (-\Delta, \Delta)$.
    We take these samples, calculate the corresponding forces, and then fit to a linear model of the forces
    \begin{equation}
        F^{e}_{\alpha} \equiv -\sum_j \Phi^{e}_{\alpha j} x_j.
    \end{equation}
    We do this by ordinary least squares regression, so in the limit that we have perfectly sampled this region, the loss function and its derivatives are
    \begin{equation}
    \label{eq:loss_function}
    \begin{split}
        L &= \int_{-\Delta}^{\Delta} d\boldsymbol{x} \left[ \sum_j (\Phi^{e}_{\alpha j} - \Phi_{\alpha j}) x_j - \frac{1}{2} \sum_{j k} \Gamma_{\alpha j k} x_j x_k - \frac{1}{6} \sum_{j k l} \Lambda_{\alpha j k l} x_j x_k x_l \right]^2\\
        \implies \frac{\partial L}{\partial \Phi^{e}_{\alpha \beta}} &= 2 \int_{-\Delta}^{\Delta} d\boldsymbol{x} \left[ \sum_j (\Phi^{e}_{\alpha j} - \Phi_{\alpha j}) x_j - \frac{1}{2} \sum_{j k} \Gamma_{\alpha j k} x_j x_k - \frac{1}{6} \sum_{j k l} \Lambda_{\alpha j k l} x_j x_k x_l \right] x_{\beta}.
    \end{split}
    \end{equation}
    where $\boldsymbol{x}$ is shorthand for all of the coordinates.
    The only nonzero terms in this integral are those in which there are even powers of x:
    \begin{equation}
    \begin{split}
        \int_{-\Delta}^{\Delta} d\boldsymbol{x} \ x_j x_{\beta} &= (2 \Delta)^{3 N} \frac{\Delta^2}{3}, \quad (j = \beta)\\
        \int_{-\Delta}^{\Delta} d\boldsymbol{x} \ x_j x_k x_l x_{\beta} &=
        \begin{cases}
            (2 \Delta)^{3N} \frac{\Delta^4}{9}, &\quad \text{(two pair)},\\
            (2 \Delta)^{3N} \frac{\Delta^4}{5}, &\quad (j = k = l = \beta)
        \end{cases}
    \end{split}
    \end{equation}
    with all other cases being zero.
    ``Two pair'' refers to the case in which there are two pairs of matching indices but they are not all the same.

    Looking at the second line of \eqref{eq:loss_function}, there is only one nonzero term in the first sum ($j = \beta$) and no nonzero terms in the second sum.
    For the last sum there is one term in which all of the indices match $(j = k = l = \beta)$ and numerous two pairs, in which one of the indices must be equal to $\beta$ and the other two must equal each other.
    These two pairs may all be grouped into one sum using the permutational invariance of $\Lambda$.
    Altogether, the derivative of the loss function is
    \begin{equation}
        \frac{1}{2} \frac{\partial L}{\partial \Phi^{e}_{\alpha \beta}} = (2 \Delta)^{3 N} \frac{\Delta^2}{3} (\Phi^{e}_{\alpha j} - \Phi_{\alpha j}) - \frac{1}{6} (2\Delta)^{3N} \Delta^4 \left[ \frac{1}{5} \Lambda_{\alpha \beta \beta \beta} + \frac{1}{3} \sum_{i \ne \beta} \Lambda_{\alpha \beta i i} \right].
    \end{equation}
    Setting this derivative equal to zero and solving for $\Phi^{e}_{\alpha \beta}$ yields
    \begin{equation}
        \Phi^{e}_{\alpha \beta} = \Phi_{\alpha \beta} + \frac{1}{2} \Delta^2 \left( \frac{1}{5} \Lambda_{\alpha \beta \beta \beta} + \frac{1}{3} \sum_{i \ne \beta} \Lambda_{\alpha \beta i i} \right).
    \end{equation}
    As in the case of the centered finite difference, including more terms in our original expansion would only change this result by introducing terms that scale as $\Delta^4$ at best.

    We see that this random displacement method yields the force constants with a systematic error that goes as $\Delta^2$, just like a centered finite difference estimate.
    In fact, the factor in front of the $\Delta^2$ may possibly be smaller in this alternate approach, depending on the sum over $\Lambda$.
    This is not cause for alarm since this approach requires more work than performing a centered finite difference.

    In practice, we can only fit to a finite number of force calculations.
    However, from this analysis we can see that the $\mathcal{O}(\Delta^2)$ behavior is preserved so long as we fit to an inversion-symmetric set of samples.
    Furthermore, as we showed in \S \ref{s:dft_tests}, even in the presence of statistical noise, this approach is practical.

    This analysis is not specific to the problem of forces in a solid, so this approach is usable in the same contexts where finite difference estimates are viable.
    The practical difference is that this method requires more calculations, but for a fixed $\Delta$ can generally yield larger numbers, if improved resolution is desired.

\end{document}